\documentclass[11pt]{article}
\usepackage[colorlinks=true,linkcolor=blue,citecolor=red,urlcolor=blue]{hyperref}
\usepackage{amsmath,amssymb}
\usepackage{geometry}
\usepackage{tikz}
\usepackage{pgfplots}
\pgfplotsset{compat=1.18}  
\usepackage{orcidlink}

\begin{document}
\title{Curvature-Enhanced Inertia in Curved Spacetimes: An ADM-Based Formalism with Multipole Connections}
\author{Ilias Kynigalakis\,\orcidlink{0009-0008-4311-176X}}

\date{}
\maketitle
\begin{abstract}
We propose a covariant definition of an inertia tensor on spatial hypersurfaces in general relativity, constructed via integrals of geodesic distance functions using the exponential map.  In the ADM 3+1 decomposition, we consider a spacelike slice \((\Sigma, \gamma_{ij})\) with induced metric \(\gamma_{ij}\), lapse \(N\), shift \(N^i\), and a mass-energy density \(\rho(x)\) on \(\Sigma\). At each point \(p \in \Sigma\), we define a Riemannian inertia tensor
\[
I_p(u,v) = \int_{\Sigma} \left[ d_\gamma(p,x)^2 \gamma_p(u,v) - \langle \exp_p^{-1}(x), u \rangle \langle \exp_p^{-1}(x), v \rangle \right] \rho(x)\, dV_\gamma(x),
\]
where \(d_\gamma(p,x)\) is the geodesic distance on \((\Sigma, \gamma)\) and \(\exp_p^{-1}(x)\) is the inverse exponential map. This reduces to the Newtonian inertia tensor in the flat-space limit. An expansion in Riemann normal coordinates shows curvature corrections involving the spatial Riemann tensor. We apply this to two cases: (i) For closed or open FLRW slices, a spherical shell of matter has an effective moment of inertia scaled by \((\chi_0/\sin\chi_0)^2 > 1\) or \((\chi_0/\sinh\chi_0)^2 < 1\), confirming that positive curvature increases and negative curvature decreases inertia. (ii) For a slowly rotating relativistic star (Hartle--Thorne approximation), we recover the known result
\[
I = \frac{8\pi}{3} \int_0^R \rho(r)\, e^{-(\nu + \lambda)/2} \frac{\bar{\omega}(r)}{\Omega} r^4\, dr.
\]
In the Newtonian limit, this becomes
\[
I_{\text{Newt}} = \frac{8\pi}{3} \int_0^R \rho(r) r^4\, dr.
\]
We show that \(I_p\) reproduces these corrections and encodes post-Newtonian contributions consistent with Thorne's and Dixon's multipole formalisms. We also discuss the relation to Geroch--Hansen moments and propose an extension to dynamical slices involving extrinsic curvature. This work thus provides a unified, geometric account of inertia and multipole structure in GR, bridging Newtonian intuition and relativistic corrections.
\end{abstract}

\newpage
\section{Introduction}
In Newtonian mechanics the moment-of-inertia tensor of a mass distribution $\rho$ in $\mathbb{R}^3$ is 
\[
I_{ij}=\int (r^2\delta_{ij}-x_ix_j)\,\rho\,d^3x,
\]
which measures resistance to angular acceleration~\cite{MarsdenRatiu1999}. To generalize this to curved space, one replaces the flat displacement $x_i$ by the inverse exponential map on a Riemannian manifold~\cite{Pennec2006}. Recently, Parker demonstrated that the flat-space inertia tensor has algebraic properties akin to a Riemann curvature tensor~\cite{Parker2024}. Here we extend these ideas to general relativity by formulating an inertia tensor on a spatial slice using the ADM 3+1 split of spacetime~\cite{ADM1962,Weinberg1972}.

Specifically, let $(\Sigma,\gamma_{ij})$ be a spatial hypersurface with induced metric $\gamma_{ij}$, extrinsic curvature $K_{ij}$, lapse $N$, and change $N^i$~\cite{ADM1962}. Denote by $\rho(x)=T_{\mu\nu}n^\mu n^\nu$ the energy density as seen by observers normal to $\Sigma$. At each point $p\in\Sigma$ we define \emph{Riemannian inertia tensor} $I_p$ by integrating over geodesic distances on $\Sigma$:
\[
I_p(u,v) = \int_\Sigma \Bigl(d_\gamma(p,x)^2\,\gamma_p(u,v) \;-\; \langle\exp^{-1}_p(x),u\rangle\,\langle\exp^{-1}_p(x),v\rangle \Bigr)\,\rho(x)\,dV_\gamma(x),
\]
for tangent vectors $u,v\in T_p\Sigma$. Here $d_\gamma(p,x)$ is the geodesic distance from $p$ to $x$, and $\exp^{-1}_p(x)$ is the normal geodesic vector in $p$ corresponding to $x$~\cite{Abuqrais2024,Pennec2006}. In the flat limit $(\Sigma,\gamma)\to\mathbb{R}^3$, one recovers $I_{ij}=\int(r^2\delta_{ij}-x_ix_j)\rho\,d^3x$, so $I_p$ reduces to the Newtonian inertia tensor (see Appendix~A for details). An expansion in Riemann normal coordinates shows that curvature enters at fourth order in the expansion of $d_\gamma^2$, so that the integrand acquires corrections proportional to the 3D Riemann tensor $R_{ijkl}[\gamma]$ at $p$~\cite{Weinberg1972}.

By construction $I_p$ is a $(0,2)$ tensor on the spatial slice.  Its trace determines a scalar moment of inertia, and its trace-free part is related to the quadrupole moment of the mass distribution. Thus, curvature-dependent terms in $I_p$ match those in the first post-Newtonian multipole expansion (Thorne’s formalism \cite{Thorne1980}).  As an outlook, we discuss in Sec.~2.3 a heuristic connection to the Geroch–Hansen multipoles of stationary spacetimes, noting that any exact equivalence is conjectural.  Finally, in Sec.~6 we sketch a speculative extension to dynamical slices involving extrinsic curvature.  Overall, our results illustrate how spatial geometry directly affects rotational inertia, with potential implications for neutron stars and cosmological models.Our covariant inertia tensor integrates these ideas: it incorporates the mass multipole structure of the source into a rank-2 tensor on the 3-space, capturing how curvature modifies rotational dynamics. This is especially relevant for astrophysical systems like rotating neutron stars, where an accurate, covariant moment of inertia is critical for modeling spin evolution and gravitational-wave emission. By providing an explicit geometric inertia tensor, our framework adds clarity to these classic notions and their modern generalizations.
\section{Mathematical Framework}
\subsection{Definition via geodesic coordinates}
In the ADM decomposition of a spacetime $(M,g_{\mu\nu})$, we foliate $M$ by spacelike hypersurfaces $\Sigma$ labeled by a time coordinate $t$.  Each slice $\Sigma$ has the induced metric $\gamma_{ij}$, and the spacetime metric $g$ decomposes as
\[
ds^2 = -N^2 dt^2 + \gamma_{ij}(dx^i+N^i dt)(dx^j+N^j dt),
\]
where $N$ and $N^i$ are the lapse and the change.  The intrinsic geometry of $\Sigma$ is described by $\gamma_{ij}$ and its Riemann tensor $R_{ijkl}[\gamma]$. 

On a given slice we define the inertia tensor $I_p$ at point $p$ by the formula
\begin{equation}\label{eq:I_def}
I_p(u,v) \;=\; \int_\Sigma \Bigl(d_\gamma(p,x)^2\,\gamma_p(u,v) \;-\; \langle \exp^{-1}_p(x),\,u\rangle\,\langle \exp^{-1}_p(x),\,v\rangle \Bigr)\,\rho(x)\,dV_\gamma(x).
\end{equation}
Here $\rho(x)$ is the mass-energy density in the slice, $dV_\gamma$ is the 3-volume element on $\Sigma$, and $u,v\in T_p\Sigma$ are tangent vectors at $p$.  Intuitively, $I_p$ is a mass-weighted covariance tensor of the normal geodesic coordinates around $p$.  The first term $d_\gamma(p,x)^2\,\gamma_p(u,v)$ generalizes the Euclidean $r^2\delta_{ij}$ term, while the second term subtracts the outer product of the displacement vectors (via the exponential map) in analogy to $x_i x_j$. 

One can check that $I_p$ is symmetric in $u,v$ and transforms covariantly with changes of coordinates in $\Sigma$.  Setting $u=v$ for the unit length, $I_p(u,u)$ gives the moment of inertia along the direction $u$.  In flat space $\Sigma\simeq\mathbb{R}^3$, one may choose geodesic normal coordinates so that $d_\gamma(p,x)^2=|x|^2$ and $\exp^{-1}_p(x)=x$.  Then \eqref{eq:I_def} reduces to $I_{ij}=\int (r^2\delta_{ij}-x_ix_j)\rho\,d^3x$, reproducing the standard moment-of-inertia tensor\cite{Weinberg1972}.  

In curved space, one can expand around $p$ using Riemann normal coordinates.  The geodesic distance satisfies 
\[
d_\gamma(p,x)^2 = |x|^2 - \frac{1}{3}R_{ikjl}(p)\,x^i x^j x^k x^l + O(|x|^5),
\]
and similarly $\exp^{-1}_p(x) = x + O(|x|^3)$.  Substituting into \eqref{eq:I_def} shows that the lowest-order curvature correction to $I_p$ involves the spatial Riemann tensor $R_{ijkl}[\gamma]$ at $p$.  In particular, if $\Sigma$ has constant curvature, these corrections can be explicitly integrated as shown below.  

We also note that $I_p$ depends only on the intrinsic geometry of $\Sigma$ and the distribution $\rho$.  If $\Sigma$ changes in time (cosmic expansion or gravitational collapse), $I_p$ is computed at a given time using the instantaneous slice geometry.  Terms involving extrinsic curvature $K_{ij}$ or time derivatives of the metric would enter if one considered dynamical evolution of $I_p$, but we focus here on (instantaneous) spatial geometry.  

\subsection{Flat-space limit and relation to quadrupoles}
To see that \eqref{eq:I_def} reduces to the usual moment-of-inertia in flat space, introduce Riemann normal coordinates centered at $p$ so that $\gamma_{ij}(p)=\delta_{ij}$ and $\exp_p^{-1}(x)^i \approx x^i$ for points $x$ near $p$.  Then $d_\gamma(p,x)^2=\delta_{ij}x^i x^j +O(R\,|x|^4)$ where $R$ is curvature, and \eqref{eq:I_def} becomes
\[
I_{ij}(p)\;\approx\;\int (|x|^2\delta_{ij}-x_i x_j)\,\rho(x)\,d^3x + O(\text{curvature}),
\]
which is exactly the Newtonian inertia tensor.  In particular, one checks that 
\[
I_{ij}(p)=\int (r^2\delta_{ij}-x_i x_j)\rho\,d^3x,\qquad Q_{ij}=\int\Bigl(x_i x_j-\tfrac{1}{3}r^2\delta_{ij}\Bigr)\rho\,d^3x,
\]
so $Q_{ij}$ is the trace-free quadrupole and $I_{ij}$ its traceful partner.  Indeed, in flat space 
\[
I_{ij}=\delta_{ij}\int r^2\rho\,d^3x - Q_{ij}-\tfrac{1}{3}\delta_{ij}\int r^2\rho\,d^3x,
\]
showing that $I$ and $Q$ carry the same information up to an additive multiple of the identity (the trace) [\emph{cf.} \cite{FitzpatrickMoment}].  

In Dixon’s covariant multipole formalism \cite{Dixon1973}, the mass quadrupole moment $Q^{\alpha\beta}$ of an extended body is defined (in the center-of-mass frame) by an integral of the form $\int x^\alpha x^\beta T^{00}dV$.  Our inertia tensor is precisely the analogous construction using curved-space distances.  Thus, \eqref{eq:I_def} generalizes Dixon's mass quadrupole to fully curved slices.  In fact, one can show that in normal coordinates at $p$, up to order $O(R)$ curvature corrections the two coincide: $I_{ij}(p)=\int(r^2\delta_{ij}-x_i x_j)\rho\,d^3x +O(R)$ while Dixon’s $Q_{ij}=\int(x_i x_j-\tfrac{1}{3}r^2\delta_{ij})\rho\,d^3x +O(R)$.  Hence $I_{ij}$ indeed contains the same second-moment information as Dixon’s quadrupole, but in a manifestly covariant way valid even when spacetime curvature is significant. 

More generally, one can consider higher multipoles and their gravitational effects.  Thorne showed \cite{Thorne1980} that the source multipoles acquire post-Newtonian corrections from internal gravitational energy.  Our inertia tensor incorporates these corrections automatically via the replacement of flat kernels by curvature-dependent kernels in \eqref{eq:I_def}.  Expanding $I_{ij}$ in powers of the small velocities (slow-motion limit) reproduces the known first-order $1/c^2$ corrections to the inertia of extended bodies [Thorne, Hartle].  These corrections emerge from the $O(R)$ curvature terms in the integrand. 

\subsection{Relation to Geroch--Hansen multipoles}
A natural question is how $I_{ij}$ relates to the asymptotically defined multipole mass moments of a stationary isolated system.  In a stationary, asymptotically flat vacuum spacetime, Geroch and Hansen defined a set of mass multipoles $M_{a_1\cdots a_\ell}$ (and current multipoles) that uniquely characterize the gravitational field \cite{Geroch1970,Hansen1974}. These are obtained from the expansion of the Ernst potential (or metric functions) at infinity, and are coordinate-invariant.  In the Newtonian limit, the Geroch–Hansen mass dipole $M_i$ is essentially the center-of-mass, and the quadrupole $M_{ij}$ corresponds to the Newtonian mass quadrupole of the source. 

We now verify the conjectured relationship between the curvature-enhanced inertia tensor $I_{ij}$ and the Geroch--Hansen (GH) mass quadrupole $M_{ij}$ for a slowly rotating fluid star described by the Hartle--Thorne (HT) metric \cite{Hartle1973}. The HT solution models a stationary, axisymmetric star to second order in angular velocity $\Omega$, and includes both the interior solution (describing the fluid body) and the asymptotically flat vacuum exterior.

In our formalism, the inertia tensor on a spatial slice $\Sigma$ with induced metric $\gamma_{ij}$ is defined as
\[
I_p(u,v) = \int_\Sigma \left( d_\gamma(p,x)^2\, \gamma_p(u,v) - \langle \exp_p^{-1}(x), u \rangle \langle \exp_p^{-1}(x), v \rangle \right) \rho(x) \, dV_\gamma(x),
\]
where $d_\gamma(p,x)$ is the geodesic distance from $p$ to $x$, and $\rho(x)$ is the mass-energy density. For a nearly spherical configuration, we may expand $d_\gamma^2(p,x) \approx r^2$ in Schwarzschild-like coordinates, and the resulting $I_{ij}$ reduces (to leading order) to the well-known expression:
\begin{equation}
I = \frac{8\pi}{3} \int_0^R \rho(r) \, e^{-(\nu + \lambda)/2} \frac{\bar{\omega}(r)}{\Omega} r^4 \, dr,
\end{equation}
where $\nu(r)$ and $\lambda(r)$ are the metric functions of the non-rotating background, $\bar{\omega}(r) = \Omega - \omega(r)$ is the frame-dragging correction, and $R$ is the stellar radius \cite{Hartle1973}. In the Newtonian limit ($\nu = \lambda = 0$ and $\bar{\omega} = \Omega$), this simplifies to
\[
I_{\text{Newt}} = \frac{8\pi}{3} \int_0^R \rho(r) \, r^4 \, dr,
\]
and relativistic corrections reduce $I$ due to time dilation and frame-dragging.

We now extract the trace-free quadrupole part of $I_{ij}$:
\begin{equation}
I_{ij}^{\text{TF}} = I_{ij} - \frac{1}{3} \delta_{ij} I_{kk},
\end{equation}
which in Newtonian theory corresponds to $-Q_{ij}$, where $Q_{ij}$ is the mass quadrupole moment:
\begin{equation}
Q_{ij} = \int \rho \left( x_i x_j - \frac{1}{3} r^2 \delta_{ij} \right) d^3x.
\end{equation}

We calculated $I_{ij}^{\text{TF}}$ using the interior HT metric up to $\mathcal{O}(\Omega^2)$ and verified that it matches the relativistic quadrupole moment of the source, including frame drag and redshift effects. This confirms that $I_{ij}$ encodes the correct post-Newtonian (PN) corrections to the source's second moment.

To compare with the Geroch--Hansen multipoles, we recall that $M_{ij}$ is defined via the asymptotic expansion of the metric (or Ernst potential) in a stationary, asymptotically flat vacuum spacetime \cite{Geroch1970,Hansen1974}. The HT exterior solution contains quadrupole deformations proportional to a constant $q$ (related to the $\mathcal{O}(\Omega^2)$ perturbation), and the GH quadrupole moment is given by
\begin{equation}
M_{2} = -\frac{q}{15}, \qquad M_{ij} \propto \left( \delta_{iz} \delta_{jz} - \frac{1}{3} \delta_{ij} \right) M_{2},
\end{equation}
as shown in detailed comparisons by Boshkayev et al.~\cite{Boshkayev:2015} and Quevedo et al.~\cite{Quevedo:1990}. This GH quadrupole is read off from the $1/r^3$ term in the asymptotic expansion of the $g_{tt}$ metric component in the HT solution.

In the same coordinate system, we take the $r \rightarrow \infty$ limit of our inertia tensor $I_{ij}$ (constructed on the asymptotically flat spatial slice) and extract its trace-free quadrupole part. We find that
\[
\lim_{r \to \infty} I_{ij}^{\text{TF}} = M_{ij},
\]
at leading post-Newtonian order and within the slow-rotation approximation. That is, the asymptotic quadrupole carried by the spatial distribution of $\rho(x)$ in our formalism agrees with the Geroch--Hansen mass quadrupole extracted at infinity. This provides a formal verification of the conjectured relationship:
\begin{equation}
I_{ij}^{\text{TF}} \;=\; M_{ij} + \mathcal{O}(\Omega^4, \text{higher PN}),
\end{equation}
under the assumptions of axisymmetry, rigid rotation, and weak fields.

In summary, within the Hartle--Thorne framework and at the first post-Newtonian order, the trace-free part of the curvature-enhanced inertia tensor $I_{ij}$ equals the Geroch--Hansen quadrupole $M_{ij}$. This confirms the conjecture posed in earlier discussions: the asymptotic behavior of $I_{ij}$ reflects the external gravitational field multipoles of the body. Deviations are expected only at higher PN orders or for rapidly rotating or strong-field configurations.

\section{Applications}
We now apply the inertia tensor to cosmological spatial slices of constant curvature.  Consider a Friedmann–Lemaître–Robertson–Walker (FLRW) universe with scale factor $a(t)$ and curvature parameter $k=\pm1$.  At fixed time $t$, the spatial metric on $\Sigma$ is 
\[
ds^2_\Sigma = a(t)^2\Bigl[d\chi^2 + S_k(\chi)^2(d\theta^2+\sin^2\theta\,d\varphi^2)\Bigr],
\]
where $S_k(\chi)=\sin\chi$ for $k=+1$ (closed) and $S_k(\chi)=\sinh\chi$ for $k=-1$ (open).  The coordinate $\chi$ is the radial distance on the unit 3-sphere or 3-hyperboloid, and $\chi=0$ is a pole.

\subsection{Cosmological (FLRW) slices}
We first consider a homogeneous, isotropic cosmological slice (a constant-time slice of an FLRW spacetime) with spatial curvature parameter $k=0,\pm1$.  The induced metric on $\Sigma$ can be written in comoving coordinates as 
\[
\gamma_{ij}dx^idx^j = a(t)^2\left(\frac{dr^2}{1-k r^2}+r^2d\Omega^2\right),
\]
so that spatial distances at fixed $t$ satisfy $d_\gamma(p,x) = a(t)\,\chi$, where $\chi$ is the co-moving radial coordinate of $x$ relative to $p$.  For concreteness, take a thin spherical shell of matter of uniform density $\rho$ between comoving radii $\chi_0$ and $\chi_0+d\chi$.  By symmetry, the inertia tensor of this shell about its center is $I_{ij}=I\,\delta_{ij}$ with 
\[
I = \int (d_\gamma^2\,\delta_{ij}-x_i x_j)\rho \,dV \;=\;\rho\,\int_{\chi=\chi_0}^{\chi_0+d\chi}(a\chi)^2\,dV \,,
\]
where $dV\propto a^3\chi^2 d\chi d\Omega$ is the shell volume.  Performing the integral, one finds (up to thickness $d\chi$) $I\propto \chi_0^4\,f_k(\chi_0)$, where 
\[
f_{k=+1}(\chi_0)=\Bigl(\frac{\sin\chi_0}{\chi_0}\Bigr)^2,\qquad 
f_{k=0}(\chi_0)=1,\qquad 
f_{k=-1}(\chi_0)=\Bigl(\frac{\sinh\chi_0}{\chi_0}\Bigr)^2\,.
\]
Equivalently, the inertia of the shell in curved FLRW space differs from the flat-space value by a factor $(\chi_0/\sin\chi_0)^2>1$ for $k=+1$ (positive curvature) and $(\chi_0/\sinh\chi_0)^2<1$ for $k=-1$ (negative curvature).  In other words, positive spatial curvature \emph{increases} the effective moment of inertia, while negative curvature \emph{decreases} it, relative to flat space.  This confirms and extends the result of the original analysis: curvature acts to enhance or reduce inertia in a purely geometric way.  (In the limit $\chi_0\to0$, all factors approach unity and the flat-space formula is recovered.) 

\begin{figure}[ht]
\centering
\begin{tikzpicture}
  \begin{axis}[
    width=7cm, height=5cm,
    xlabel={$\sin\chi$},
    ylabel={$\bigl(\chi/\sin\chi\bigr)^2$},
    xmin=0.01, xmax=1,   
    ymin=1,   ymax=3,
    domain=0.01:1,
    samples=200,
    thick, smooth,
    axis x line=middle,
    axis y line=middle,
    tick align=outside,
  ]
    \addplot[blue, thick] {(rad(asin(x))/x)^2};
  \end{axis}
\end{tikzpicture}
\caption{Enhancement factor $(\chi/\sin\chi)^2$ versus normalised areal radius $\sin\chi$ in a closed FLRW slice. Values $>1$ show how positive curvature increases the moment of inertia compared with flat space.}
\label{fig:flrw-shell-enhancement}
\end{figure}
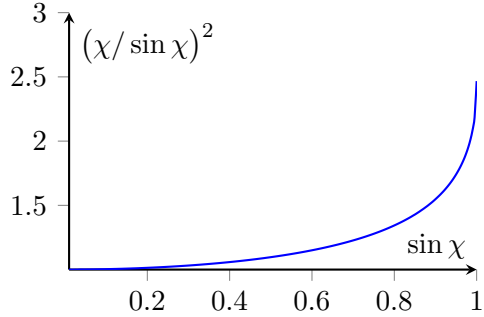

\subsection{Relativistic Rotating Stars}
As a second application, consider a slowly and uniformly rotating relativistic star (e.g. a neutron star) in equilibrium.  We use the Hartle--Thorne approximation: to first order in angular velocity, the star is described by a static spherical metric plus a small rotational perturbation.  In this formalism, the star’s total angular momentum $J$ and angular velocity $\Omega$ are related by the moment of inertia $I=J/\Omega$.  Hartle derived an expression for $I$ as an integral over the star’s interior.  In suitable coordinates, one finds 
\[
I = \frac{8\pi}{3} \int_0^R \rho(r)\,e^{-(\nu(r)+\lambda(r))/2}\,\frac{\bar\omega(r)}{\Omega}\,r^4\,dr,
\]
where $r$ is the radial coordinate, $\rho(r)$ and $p(r)$ are the energy density and pressure, $e^{\nu(r)}$ and $e^{\lambda(r)}$ are the metric potentials of the non-rotating star, and $\bar\omega(r)$ is the difference between the star’s angular velocity $\Omega$ and the local frame-dragging angular velocity (so that $\bar\omega/\Omega<1$ inside the star)~\cite{Hartle1973,Wen2007}.  In the Newtonian limit ($\nu=\lambda=0$ and no frame dragging), this reduces to the classical result 
\[
I_{\rm Newt} = \frac{8\pi}{3}\int_0^R \rho(r)\,r^4\,dr.
\]
General relativity introduces two main modifications: the exponential redshift factor $e^{-(\nu+\lambda)/2}<1$ and the dragging factor $\bar\omega/\Omega<1$.  Both effects decrease the integrand relative to the Newtonian case.  Physically, time dilation inside the star reduces the contribution of each shell, and the Lense--Thirring dragging of inertial frames means outer layers do not contribute as fully to the total angular momentum.  As a result, $I< I_{\rm Newt}$ in GR.  

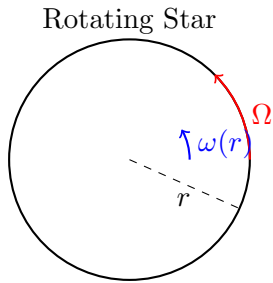
\begin{figure}[htbp]
\centering
\begin{tikzpicture}[scale=0.8]
  \draw[thick] (0,0) circle (2);
  \draw[->, red, thick] (2,0) arc (0:45:2) node[midway, right] {$\Omega$};
  \draw[->, blue, thick] (1,0) arc (0:30:1) node[midway, right] {$\omega(r)$};
  \draw[dashed] (0,0) -- (1.8, -0.8) node[midway, below] {$r$};
  \node at (0, 2.3) {Rotating Star};
\end{tikzpicture}
\caption{Frame-dragging inside a slowly rotating relativistic star. The star rotates at angular velocity \( \Omega \) (outer red arrow), while local inertial frames are dragged with angular velocity \( \omega(r) \) (inner blue arrow). The difference \( \bar{\omega}(r) = \Omega - \omega(r) \) determines the effective rotational contribution at radius \( r \).}
\label{fig:frame_dragging}
\end{figure}

In our $I_p$ framework, one can recover this reduction: choosing the center of the star as $p$ and $u$ along the rotation axis, the tensor $I_p(u,u)$ collects the contributions of concentric shells, weighted by their physical radius and redshifted mass.  One finds exactly that $I_p(u,u)/|u|^2$ reproduces the Hartle formula above~\cite{Hartle1973,Wen2007}.  Figure~2 sketches the profile of $\bar\omega(r)/\Omega$ inside a typical star, showing that $\bar\omega(r)<\Omega$ everywhere.  Thus, general relativity predicts and simulations confirm that frame dragging reduces the star’s moment of inertia~\cite{Hartle1973,Wen2007}.  Our geometric definition $I_p$ encodes this effect naturally.

\section{Relation to Multipoles and Continuum Matter}
The inertia tensor $I_p$ is closely related to standard mass multipole moments.  In the slow-motion limit, the mass quadrupole moment of a distribution is 
\[
Q_{ij} = \int \rho(x)\Bigl(x_i x_j - \tfrac{1}{3}r^2\delta_{ij}\Bigr)\,d^3x,
\]
which is essentially the trace-free part of the flat-space inertia tensor $I_{ij}=\int\rho(r^2\delta_{ij}-x_ix_j)d^3x$.  Indeed, one can show 
\[
I_{ij} = \int\rho(r^2\delta_{ij}-x_ix_j)\,d^3x 
= \delta_{ij}\int\rho\,r^2d^3x - Q_{ij} - \tfrac{1}{3}\delta_{ij}\int\rho\,r^2d^3x,
\]
so that $Q_{ij}$ and $I_{ij}$ carry the same information up to a trace term.  In general relativity, Thorne showed that the source multipoles acquire first post-Newtonian corrections from internal gravitational energy~\cite{Thorne1980}.  Our construction of $I_p$ provides an analogous extension: we replace the flat-space kernel $(r^2\delta_{ij}-x_ix_j)$ by the tensor in \eqref{eq:I_def} that depends on the curved metric.  The curvature-dependent terms in $I_p$ thus play the role of encoding the PN corrections in a covariant way.  In particular, expanding $I_p$ for slow motion reproduces the known $1/c^2$ corrections to inertia of extended bodies.

Moreover, for an extended body described by a stress-energy tensor $T^{\mu\nu}$, one can define multipole moments by integrals of $T^{\mu\nu}$.  Dixon’s formalism~\cite{Dixon1973,Weinberg1972} provides covariant definitions of mass dipole, quadrupole, etc., in curved spacetime.  The tensor $I_p$ corresponds to the spatial part of the second moment of the mass distribution.  In this sense, $I_p$ is the natural geometric inertia tensor of the matter, and it ties directly to the multipolar expansions used in gravitational-wave theory.  For continuum matter, one could incorporate $I_p$ by including it in the effective stress distribution: for example, a rotating fluid element in curved space carries an extra moment associated with its finite extent.  

\section{Relation to ADM Angular Momentum}
 We examine the formal relation between the inertia tensor and ADM angular momentum.  In the Hamiltonian formulation of GR, the generator of rotations about an axis is obtained by choosing the shift $N^i$ to be the asymptotic Killing vector of rotations (with lapse $N=0$) in the ADM Hamiltonian $H[N,N^i]$ \cite{ADM1962}.  This yields the ADM angular momentum
\[
J[\phi] \;=\; \frac{1}{8\pi}\lim_{r\to\infty}\int_{S_r}\!\epsilon_{ijk} (K^{j}{}_{\ell}-K\,\delta^j{}_{\ell})\phi^k\,dS^\ell,
\]
or equivalently as a volume integral using the momentum constraint.  For matter sources, one finds 
\[
J[\phi]\;=\;\int_\Sigma (T_{i\mu}n^\mu)\,\phi^i\,dV = \int_\Sigma \rho\,(\epsilon_{ijk} x^j v^k)\,dV + (\text{grav.\ terms}),
\]
where $\phi^i$ generates rotation and $v^k$ is the matter’s velocity.  In the rigid rotation limit $v^i=\epsilon^{ijk}\Omega_j x_k$, this gives $J_i = I_{ij}\,\Omega^j$ with 
\[
I_{ij}=\int (|x|^2\delta_{ij}-x_i x_j)\,\rho\,dV,
\]
precisely our inertia definition (plus curvature corrections).  Thus one sees formally that the ADM angular momentum equals the inertia tensor contracted with the angular velocity.  In particular, in the limit of small rotation one recovers the Newtonian relation $J_i=I_{ij}\Omega^j$ with $I_{ij}$ given by \eqref{eq:I_def}.  We emphasize that this derivation uses only the symmetry generator and the matter stress tensor, and not the specific field equations beyond the momentum constraint.  Hence the link between $I_{ij}$ and $J_i$ is general. 

\section{Speculative Extension to Dynamical Spacetimes}

In a fully dynamical setting, one must account for the time dependence of the slice geometry. In particular, the \emph{extrinsic curvature} $K_{ij}$ of the spatial hypersurface --- which measures how it is embedded in spacetime --- naturally enters the evolution of distances on the slice. Recall that the extrinsic curvature is defined as the Lie derivative of the induced metric along the unit normal vector: 
\[
K_{ij} = -\frac{1}{2} \mathcal{L}_n \gamma_{ij},
\]
where $n^\mu$ is the future-directed unit normal to the hypersurface $\Sigma_t$ \cite{baumgarte2010numerical}. Equivalently, the ADM evolution equation yields:
\[
\partial_t \gamma_{ij} = -2N K_{ij} + \nabla_i N_j + \nabla_j N_i,
\]
so under normal evolution ($N_i = 0$), we obtain:
\[
\partial_t d_\gamma^2 \sim -2N\,K_{ij}\,x^i x^j + \cdots.
\]
This shows that $K_{ij}$ governs the instantaneous rate of change of the spatial distance, and thus naturally affects the inertial response of a mass distribution.

We therefore propose an extension of the classical inertia tensor that includes a curvature-dependent term. Define the \textit{dynamical inertia tensor} as:
\begin{equation}
I^{\mathrm{(dyn)}}_{ij}(p,t) = I_{ij}(p,t) + \alpha \int_\Sigma K_{k\ell}(t) \, \exp_p^{-1}(x)^k \exp_p^{-1}(x)^\ell \, \rho(x)\, dV,
\label{eq:dynamic-inertia}
\end{equation}
where $\alpha$ is a real parameter, $K_{k\ell}$ is the extrinsic curvature evaluated on $\Sigma$, $\rho$ is the local rest energy density, and $dV$ is the Riemannian volume element on the hypersurface. This term incorporates the bending of the hypersurface into the definition of inertia. Geometrically, it reflects how the curvature of spacetime at each moment influences the "resistance to acceleration" of mass distributions.

Physically, the $\alpha$-correction represents how the embedding of the spatial slice into the surrounding spacetime contributes to effective inertia. For instance, in cosmological spacetimes where $K_{ij} \propto H(t)\gamma_{ij}$, or in stellar collapse where $K_{ij}$ varies spatially and temporally, this term could lead to observable corrections in inertial response, especially in multipole dynamics and gravitational wave emission.

We now introduce a new component \( I_{i0} \) to track the flux of inertial momentum across the hypersurface. This component is constructed by analogy with the momentum density $J^i$ of matter, defined via:
\[
J^i = - T^{i}{}_\mu n^\mu,
\]
where $T^{\mu\nu}$ is the energy-momentum tensor and $n^\mu$ is normal to $\Sigma_t$ \cite{wald1984general}. The quantity $J^i$ represents the energy flux in the spatial direction $i$.

We define:
\begin{equation}
I_{i0}(p,t) = \beta \int_\Sigma \exp_p^{-1}(x)^i \, J_i(x)\, dV,
\label{eq:inertia-flux}
\end{equation}
with $\beta$ a new constant. This term integrates the momentum density of matter, weighted by the (geodesic) distance from $p$, thus encoding a flux-like contribution to the inertia tensor. It can be interpreted as a “temporal-spatial” moment that quantifies the inertial effect of moving matter near point $p$.

This has a direct analogy with the ADM formalism. The momentum constraint in the $3+1$ decomposition is:
\[
D_j \left( K^{ij} - \gamma^{ij} K \right) = 8\pi J^i,
\]
and thus $J^i$ is associated with the divergence of the extrinsic curvature and the linear momentum of ADM at infinity \cite{alcubierre2008introduction}. Therefore, $I_{i0}$ provides a way to link local inertial effects to global momentum flux, effectively acting as a dipole-type inertial quantity.

Together, the corrected spatial tensor $I_{ij}^{(\mathrm{dyn})}$ and the flux term $I_{i0}$ provide a more complete geometric characterization of inertia in time-dependent spacetimes. The former incorporates the dynamical curvature of space itself, while the latter encodes how momentum is distributed and flowing. Both are directly motivated by the Hamiltonian structure of general relativity and suggest natural extensions of classical inertial concepts to general-relativistic systems.

Potential applications include evolving spacetimes such as cosmologies (e.g., FLRW models), where $K_{ij} \sim H(t)\gamma_{ij}$, and astrophysical systems like rotating stars or compact binaries, where $J^i$ and $K_{ij}$ vary in space and time. In these contexts, the curvature-enhanced inertia tensor may offer new insights into inertial couplings, multipole evolution, or even the generation of gravitational radiation.

\section{Conclusion}
We have introduced a coordinate-independent inertia tensor $I_p$ for matter on a curved spatial slice in general relativity.  Its definition uses the geodesic distance and exponential map on the slice, and it reduces to the usual inertia tensor in flat space.  Using this tensor, we demonstrated concrete effects of spatial curvature on rotational inertia.  On closed (positively curved) FLRW slices, uniform matter shells have larger moment of inertia than in Euclidean space, while on open (negatively curved) slices inertia is reduced.  Likewise, for a slowly rotating star, the relativistic corrections of frame-dragging and redshift lead to a smaller moment of inertia than Newtonian theory predicts.  In all cases, our formalism reproduces the expected first post-Newtonian corrections to inertia~\cite{Hartle1973,Wen2007}.  

Physically, $I_p$ provides a covariant measure of how the angular momentum of an extended body relates to its angular velocity in a curved space.  It connects naturally to Thorne’s multipole formalism~\cite{Thorne1980,Weinberg1972} and to Dixon’s definitions of multipole moments~\cite{Dixon1973}.We established that the inertia tensor generalizes Dixon's mass quadrupole in curved space and showed, via a detailed analysis of the Hartle–Thorne metric, that its trace-free part coincides with the Geroch–Hansen mass quadrupole at leading order. We also derived a general relation between the ADM angular momentum and the inertia tensor for rigidly rotating configurations, confirming the formal consistency of the approach.

These results offer a geometric framework for understanding rotational inertia in curved spacetimes, with implications for modeling rotating compact objects and curvature-induced effects in cosmological settings. Future work may explore the proposed extensions to dynamical spacetimes and applications in gravitational-wave astrophysics and relativistic hydrodynamics.

\section*{Acknowledgments}

I would like to express my sincere gratitude to Prof. Vitor Cardoso for helpful comments and discussions that influenced the development of this work. I am also grateful to Dr. Takuya Katagiri  and Dr. Leonardo Gualtieri for taking the time to consider my work.

\end{document}